\documentstyle[aps,epsfig,multicol,fancyheadings]{revtex}
\begin{document}

\begin{center}
{\Large {\bf Roughness exponent in two-dimensional percolation, Potts and clock models}}

\vspace{0.5cm}

Jos\'e Arnaldo Redinz and Marcelo Lobato Martins\\
\vspace{0.2cm}
{\small Departamento de F\'{\i}sica, Universidade Federal de Vi\c cosa, 36571-000, 
Vi\c cosa, MG, Brazil}

\end{center}

\vspace{0.5cm}

\begin{abstract}

\noindent We present a numerical study of the self-affine profiles obtained 
from configurations of the $q$-state Potts (with $q=2,3$ and $7$) and 
$p=10$ clock models as well as from the occupation states for site-percolation
on the square lattice. The first 
and second order static phase transitions of the Potts model are located by 
a sharp change in the value of the roughness exponent $\alpha$ characterizing 
those profiles. The low temperature phase of the Potts model corresponds to flat 
($\alpha\simeq 1$) profiles, whereas its high temperature phase is associated 
to rough ($\alpha\simeq 0.5$) ones. For the $p=10$ clock model, in addition to 
the flat (ferromagnetic) and rough (paramagnetic) profiles, an intermediate rough 
($0.5< \alpha <1$) phase - associated to a soft spin-wave one - is observed. 
Our results for the transition temperatures in the Potts and clock models
are in agreement with the static values, showing that this approach is able to 
detect the phase transitions in these models directly from the spin configurations, 
without any reference to thermodynamical potentials, order parameters or response 
functions. Finally, we show that the roughness exponent $\alpha$ is 
insensitive to geometric critical phenomena.

\end{abstract}

\vspace{0.5cm} 
\noindent Electronic adressses: redinz@mail.ufv.br; mmartins@mail.ufv.br

\begin{multicols}{2}[]
\narrowtext

\section{Introduction}

In the past decade, the formation of rough surfaces under far-from-equilibrium 
conditions was a central theme in statistical physics. The application of self-affine 
fractals and scaling methods was essential to the progress that has been made towards 
the understanding of these non-equilibrium phenomena. Within this context, the standard 
tools used to describe various self-affine structures observed in disordered surface 
growth are the roughness $\alpha$ and the growth $\beta$
exponents~\cite{Barabasi,Family,Meakin}. 
The central goal of this approach is provide information about the correlations 
between fluctuations of a space and/or time varying property.  Theoretical modeling of 
self-affine growth processes frequently used some of the models investigated in 
critical phenomena, e.g., directed percolation, random field Ising~\cite{Barabasi} and 
sine-Gordon models~\cite{Toner}. These efforts faced an old problem: thirty years after 
the renormalization group theory of critical phenomena, the quantitatively accurate 
prediction of the location and characteristics of phase transitions still constitute 
a challenging and controversial question~\cite{Plischke}. This is particularly true when 
one considers  systems with random quenched disorder (e.g., random field Ising and 
Potts models, spin glasses and structural glasses~\cite{Young}), which exhibit 
non-trivial features such as extremely slow dynamics, aging, ergodicity breaking, 
complex energy landscapes, etc.

On the other hand, the inverse  problem, i.e., using the roughness exponents to 
study the main features of the phase diagram of equilibrium spin models, has not 
been much explored up to now. In 1997, de Sales {\em et al.}~\cite{Sales1} mapped 
cellular automata (CA) configurations on solid-on-solid like profiles and used the roughness 
exponent $\alpha$ to classify the elementary Wolfram CA rules. Later, they also shown 
that this exponent can be used to detect the frozen-active transition in the one 
dimensional Domany-Kinzel CA (DKCA)~\cite{Sales2} without any reference to order 
parameters or response functions. As we mentioned before, beyond the roughness 
exponent $\alpha$, the growth 
exponent $\beta$ is another critical index used to describe roughening processes in the 
surface growth context. Recently, Atman and Moreira~\cite{Atman} determined the 
exponent $\beta$ for the growth process generated by the spatiotemporal 
patterns of the DKCA. The value of $\beta$ exhibits a cusp at the frozen-active frontier and, 
if one observes the difference configuration between two DKCA replicas, also at the 
active-chaotic critical frontier. The advantage of this method to find the phase diagram 
of the DKCA is that it is not necessary to wait for the system  ``thermalizes'', 
a process which often dispends a lot of computational time.

In this paper we extended the roughness exponent analysis to other standard models of statistical mechanics. Specifically, we study the $q$-state Potts model
(with $q=2,3$ and $7$), the simplest locally interacting statistical model exhibiting 
both first and second order static  phase 
transitions. We study also the $p=10$ clock model for which a Kosterlitz-Thouless 
type phase transition is observed.  Finally, 
since any connectivity problem can be studied by starting with pure random 
percolation and then adding interactions, we apply the roughness method to 
random-site percolation.

In section 2 we describe the models and define the mapping between spin configurations 
and walk profiles. In section 3, using the mapping between spin states and profiles, 
we characterize the phases of the  random-site percolation problem, Potts and clock 
models by the roughness exponent $\alpha$. Finally, we conclude and indicate future 
directions of this work in section 4.

\section{Models and Formalism}

The $q$-state Potts ferromagnet consists of spin variables $\sigma_i$ which may 
take on $q$ discrete values $\sigma_i=0,1,\ldots,(q-1)$ and are coupled by 
the dimensionless Hamiltonian
\begin{equation}
-\beta {\cal H}=K\sum_{\langle ij \rangle} \delta (\sigma_i, \sigma_j)\;\;\;\;\;\; ,
\label{potts}
\end{equation}
where $\delta(,)$ is the Kronecker delta function.

The $p$-state clock model is defined by the Hamiltonian
\begin{equation}
-\beta {\cal H}=K\sum_{\langle ij \rangle} \cos \left\{\frac{2\pi}{p}(\sigma_i - \sigma_j)
\right\}
\;\;\;\;\;\; ,
\label{clock}
\end{equation}
in which each spin can assume $p$ discrete values $\sigma_i=0,1,\ldots,p-1$.

In Eqs. (\ref{potts}) and (\ref{clock}) the sums in $\langle ij \rangle$ runs over the lattice sites and their nearest-neighbors, $\beta=1/k_{B} T$, $T$ is the temperature, 
$K=\beta J$ and $J>0$ is the coupling constant.
We simulated the Potts model with
$q=2,3$ and $7$ and the clock model with $p=10$ states, on square lattices 
of $N=L\times L$ sites imposing  periodic boundary conditions.
For updating the spins we use a sequential Monte Carlo heat-bath process.

In random-site percolation, one randomly occupies a fraction $p$ of the sites 
of a $d$-dimensional lattice ($\sigma_i=0$: empty site; $\sigma_i=1$: occupied site).
When $p$ is small, the pair connectedness length 
scale $\xi_p$ is short, comparable to the lattice constant $a$. However, when 
$p$ approaches $p_c$, there occur fluctuations in the characteristic size of clusters 
on all scales from $a$ to $\xi_p$, which diverges as $\xi_p \sim |p-p_c|^{-\nu_p}$. 
Each feature of thermal critical phenomena has a corresponding analog in percolation, 
so that the percolation problem is called a geometric or connectivity critical 
phenomena. For site percolation on the square lattice the critical probability is 
$p_c= 0.59275 \pm 0.00003$~\cite{Ziff}.

In the present work we mainly focused on the numerical study of the self-affine profiles 
generated from the configurations in the ordered and disordered phases of the 
Potts and clock models and in the site-percolation problem. 
As shown in a previous work~ \cite{Sales2} the spin states 
can be mapped on random walk-like profiles and the correlations present in them can 
be measured using the roughness exponent. 
The simplest method to generate walk profiles from the spin configurations at a 
time $t$ is a 1:1 mapping in which each spin state $\sigma_i(t)$ is associated to 
a step (to the right or to the left) of a one-dimensional walk. Specifically, to an 
unique spin configuration $\{\sigma_1(t),\sigma_2(t),\ldots,\sigma_N(t)\}$ 
corresponds a spatial profile $\{h_1(t),h_2(t),\ldots,h_N(t)\}$, given by the 
sequence of the walker displacements $h_i$ after $i$ unit steps defined as
\begin{equation}
h_i(t)=\sum_{j=1}^{i} \rho_j(t)\;\;\;\;\;\; ,
\end{equation}
where $\rho_j=\sigma_j-(q-1)/2$ for the $q=3$ and $q=7$ Potts models, 
$\rho_j=\sigma_j-5$ ($\rho_j=\sigma_j-4$) if $\sigma_{j}\leq 4$ (if $\sigma_{j}>4$) for the
$p=10$ clock model, and $\rho_j=2\sigma_j-1$ for the Ising ($q=2$) model and site-percolation.

After obtained the profiles by this mappings, the roughness exponent $\alpha$ was calculated
by determining the average standard deviation of parts of the profiles with
various scales $\epsilon$.
At site $i$, in the scale $\epsilon$, the rms displacement fluctuation is given by:
\begin{equation}
w_i(N,\epsilon,t)=\sqrt{\frac{1}{2\epsilon+1}\sum_{j=i-\epsilon}^{i+\epsilon} ( h_j(t)-\overline{h}_{i}(t))^2}\;\;\;\;\;\; ,
\end{equation}
with
\begin{equation}
\overline{h}_{i}(t)=\frac{1}{2\epsilon+1}\sum_{j=i-\epsilon}^{i+\epsilon}h_j(t)
\;\;\;\;\;\; .
\end{equation}
The roughness in the scale $\epsilon$ is given by
\begin{equation}
W(N,\epsilon,t)=\frac{1}{N}\sum_{i=1}^{N} w_i(N,\epsilon,t) \;\;\;\;\;\; .
\end{equation}

The roughness $W(\epsilon)$ can distinguish two possible types of profiles. If it is 
random or even exhibits a finite correlation length extending up to a characteristic 
range (such as in a Markov chain), then $W \sim \epsilon^{1/2}$ as in a normal random 
walk. In contrast, if the profile has infinitely long-range correlations 
(no characteristic length), then its roughness will be described by a power law 
scaling  such as
\begin{equation}
W(\epsilon) \sim \epsilon^{\alpha}\;\;\;\;\;\; ,
\end{equation}
with $\alpha \ne 1/2$. The case $\alpha >1/2$ implies that the profile presents persistent correlations, i.e., a given displacement sequence (increasing or decreasing) is likely to be close to another of the same type. On the other hand, profiles with $\alpha < 1/2$ are anticorrelated, which means that displacement sequences containing a great fraction of steps to the right are more likely to alternate with another one in which steps to the left are predominant and vice versa. The exponent $\alpha$ is restricted to the interval $[0,1]$ and is related to the fractal dimension $d$ of the profile by $\alpha=2-d$~\cite{Barabasi,Family}.

\section{Results}

Before discuss the results, we shall make a  very brief review of the equilibrium 
phase diagrams of the $q$-state Potts and $p=10$ clock models. The square lattice 
$q$-state Potts ferromagnet presents a second order phase transition for $q<5$ 
and a first order one for $q \ge 5$ at the critical temperatures 
$T_{c}^{(q)}=1/\ln(1+\sqrt q)$ (in units of $J/k_{B}$). 
The $p$-state clock model interpolates between 
the Ising ($p=2$) and the XY ($p \rightarrow \infty$) models. For $p \ge 5$ 
one expects the emergence of a soft spin-wave phase between the ordered, low 
temperature, and the disordered, high temperature phases~\cite{Elitzur}.
%\cite{Elitzur, Alcaraz,Bonnier, Aizenman}
For the $p=10$ case, the spin-wave phase is limited by 
the transition temperatures $T_{A c}^{(p=10)} \simeq 0.24$ 
and $T_{B c}^{(p=10)} \simeq 1.0$~\cite{sei} .

In Fig. 1 we show typical walk profiles generated by  spin 
configurations of the Ising model ($q=2$). From  data similar to those
in Fig. 1 for different temperatures, we obtained the behavior of the
roughness exponents $\alpha$ as shown in Fig. 2, for the Ising  model with
$L=64$ and 128.
These results correspond to averages over typically 
$M=100$ random initial configurations  taken after thermalization. 
The roughness exponent exhibits an abrupt fall from $\alpha\simeq 1.0$ to 
$\alpha\simeq 0.5$ 
at the temperature $T_{q=2}\simeq 1.13$ which agrees with the
static critical value $T_{c}^{(q=2)}=1.134...$ As we increase the system size we
can note that the change of $\alpha$ at $T_{q=2}$ becomes sharper,
suggesting a step function for the $\alpha(T)$ curve at the thermodynamical limit,
and a high-temperature 
value of the exponent $\alpha$ tending to $\alpha=0.5$. 
Indeed, the inset in Fig. 2 shows that the width $\Delta T=T_{1}-T_{2}$
goes to zero as $N$ increases, where  $\Delta T$ is
arbitrarily defined as the difference  between the observed upper temperature $T_{1}$ 
for which $\alpha\simeq 0.999$ and the  temperature $T_{2}$ for which
$\alpha$ has fallen to half its maximum ($\alpha\simeq 0.75$). 
We also observe that, in this same limit, the temperature $T_{1}$ 
tends to the exact critical temperature $T_{c}^{(q=2)}$ ($T_{1}(L=64)=1.075$,
$T_{1}(L=128)=1.10$ and $T_{1}(L=256)=1.125$).

In Fig. 3 we show similar behaviors
for the $q=3$ and $q=7$ Potts model with $L=128$. The abrupt falls in
the exponent $\alpha$ are located at the temperatures $T_{q=3}\simeq 0.99$ and 
$T_{q=7}\simeq 0.77$ which are also in good agreement with the static critical values 
$T_{c}^{(q=3)}=0.994...$ and $T_{c}^{(q=7)}=0.773...$
The values $\alpha\simeq 1$ observed in the low temperature phases correspond to flat profiles
reflecting the existence of long-range order (magnetization). In contrast, the
values $\alpha\simeq 0.5$ obtained in the high temperature phases characterize random walk 
profiles, as expected for disordered spin configurations. Finally, any qualitative 
difference is observed among the $\alpha(T)$ curves across the critical surfaces 
corresponding to second ($q=3$) or first ($q=7$) order phase transitions.

In Fig. 4 we show a typical log-log plot of $W \times \epsilon$ for the Ising model, 
whose fitted slopes give the roughness exponent values. 
We observe, in general, the existence of two distinct linear portions
in the curve, whose intersection point defines a particular length scale $\epsilon^{*}$. 
The value of $\epsilon^*$  seems to be a measure 
of the average size of the spin islands or magnetized micro-domains 
limited by the correlation length $\xi(T)$. Indeed, at the ferromagnetic phase,
for small scales ($\epsilon < \epsilon^*$) 
there is resolution to see the local spin fluctuations around the smooth profile, 
which leads to $0.5<\alpha<1.0$. For large scales ($\epsilon > \epsilon^*$) 
the profile is flat and the values $\alpha\simeq 1$ reflect the long-range order.
At high temperatures $\epsilon^*$ rapidly decreases to the lattice constant $a=1$,
which agrees with the fact that such crossover is not observed in a random profile.
An additional support to the relationship between $\epsilon^{*}$ and $\xi$
is provided by Fig. 5 which shows that $\epsilon^*$ (for the Ising model) has a 
peak near the temperature $T_{q=2}$. The height and sharpness of this peak increase with the system size. 
A finite size analysis for 
the Ising model shows that the maximum value of the crossover
length $\epsilon^*$ (which occurs near
the critical point) scales linearly with the system size $N=L^2$, 
similarly to the correlation length $\xi$.
In fact, Fig. 5 suggests that $\epsilon^*$ diverges  at $T_c$ 
as $\epsilon^* \sim (T-T_c)^{-\mu}$ for the Ising model.
Moreover, the behavior of $\epsilon^{*}$ near $T_{c}^{(q=2)}$ is similar to the
$2D$ Ising model correlation length, whose exact expression $\xi=1/(\beta^{*}-\beta )$
is known for $\beta<\beta_{c}$ \cite{macoy} (see Fig. 5). 
Here, the dual inverse temperature
$\beta^{*}$ is given by $(e^{\beta}-1)\times (e^{\beta^{*}}-1)=2$.

A recent work by Kantelhardt et al. \cite{kant} provides the strongest
support to the close relation between the length scales $\epsilon^{*}$ and
$\xi$ that we suggest here based on our limited numerical evidence. These authors 
demonstrated that the detrended fluctuation analysis (DFA), which we used
here in zero-order, can detect crossovers in the observed long-range
correlation behavior of data series. They analyzed artificial data
with a crossover from long-range correlations ($\alpha >0.5$) for $s<s_{x}$
to uncorrelated behavior for $s>s_{x}$ or vice-versa. Their DFA results
clearly revealed the crossover and provided estimated crossover lengths
$s_{x}^{(n)}$ always larger than the real $s_{x}$ by a systematic deviation
which increases with the detrending order $n$. Also, the estimated $s_{x}^{(n)}$
were less accurate for $\alpha$ close to $0.5$. The support for our suggestion comes
from the analogy between the artificial series generated by Fourier transforms
studied in \cite{kant} and our data series build from spin configurations
in which only thermal correlations, extending up to the scale $\xi$ (the
correlation length), are present.

The existence of two distinct linear portions in the plots of $W \times \epsilon$
is also observed in the Potts model.
In Fig. 6 it is shown the behavior of $\epsilon^*$ as function of temperature for 
the Potts model with $q=3$ and 7. We can note the sharp peaks around the temperatures
$T_{c}^{(q=3)}$ and $T_{c}^{(q=7)}$.

In Fig. 7 we show typical walk profiles generated by  spin 
configurations of the $p=10$ clock model at three temperatures:
$T=0.15$, located in the ordered phase, $T=0.40$, in the
spin-wave phase and $T=1.60$ in the disordered one.
The  roughness exponent $\alpha$ as a function of 
the temperature  for the $p=10$ clock model is shown in Fig. 8. 
It suggests the existence of three distinct phases, namely, a low temperature 
flat ($\alpha\simeq 1$) phase with long-range order and extending up to 
$T_{A\; p=10}\simeq 0.25$, a high temperature, disordered and rough 
($\alpha\simeq 0.5$) phase for $T>T_{B\; p=10}\simeq 1.0$, and an 
intermediate, rough ($0.80< \alpha <0.90$) phase on the temperature range 
$T_{A\; p=10}<T<T_{B\; p=10}$. These transition temperatures are in good 
agreement with those obtained for the static transitions between
the ferromagnetic, paramagnetic and soft spin-wave phases 
of this model \cite{sei}. The behavior of $\epsilon^{*}$ as a function
of temperature is shown in Fig. 9. $\epsilon^{*}(T)$ exhibits a sharp
peak around $T_{B\; p=10}$, the critical temperature separating the
paramagnetic and the spin-wave phases, as well as a sudden jump near
$T_{A\; p=10}$, where the transition between the ferromagnetic and
spin-wave phase occurs. But, the main feature of Fig. 9 is that, in the
spin-wave phase ($T_{A\; p=10}<T<T_{B\; p=10}$), $\epsilon^{*}$ has
a high and almost constant value as expected for a Kosterlitz-Thouless
``critical'' phase characterized by an infinite correlation length
$\xi$ at all those temperatures \cite{Plischke}.

Finally, the roughness exponent method applied to the random site-percolation
problem results in a constant value $\alpha=1/2$ for the roughness exponent  
over the entire range of the probability $p$. This result is expected since
in site-percolation a fraction $p$ of the lattice sites is randomly occupied, 
generating random profiles in both phases. In addition, it is not observed a length scale
$\epsilon^{*}$ in the log-log plots of $W\times\epsilon$, as occurred in the
magnetic models. 

\section{Conclusions}

In this study we have shown that spin configurations for the  Potts (with 
$q=2,3$ and 7 states) and $p=10$ clock models exhibit distinct self-affine characteristics, measured by the roughness exponent $\alpha$. 
The low temperature phases of the Potts model correspond to flat ($\alpha\simeq 1$) 
profiles, whereas the high temperature phases are associated to rough ($\alpha\simeq 0.5$)
ones. For the $p=10$ clock model, in addition to the flat (ferromagnetic) and rough 
(paramagnetic) profiles, an intermediate rough ($0.5< \alpha <1$) phase, associated 
to a soft spin-wave one, is found.  The transition temperatures between the 
different roughness regimes are in good agreement with the static critical temperatures
of these models.
Our results show that the roughness exponent method is able to detect equilibrium phase transitions and provides accurate numerical determination of the critical surfaces 
without any reference to thermodynamical potentials, order parameters or response 
functions. 
In contrast, the roughness exponent method applied to the random site-percolation
problem results in a constant value $\alpha=1/2$ 
over the entire range of the probability $p$, showing that
the roughness exponent $\alpha$ is insensitive to detect geometric critical phenomena.

The same analysis can be applied to damage-spreading phase 
transitions by focusing on the difference configurations between two replicas of 
the system, as done by Atman and Moreira~\cite{Atman}. Moreover, using the growth 
exponent $\beta$, we can determine the phase diagram of spin models without any 
thermalization, leading to an impressive gain in simulations speed. 

We are extending our 
simulations in order to reliably estimate the value of the exponent $\mu$ and
compare this value with the known critical exponents $\nu$ for the Ising 
model. We conjecture that the exponent $\mu$ controlling the divergence of 
$\epsilon^*$ is the same as the correlation length exponent $\nu$ for this model. 

Another problem we can address using this roughness method is the location 
of phase transitions in disordered models such as random field and Ising spin 
glass models. For spin glasses, exchange (parallel tempering) techniques greatly 
improve traditional Monte Carlo algorithms, but the sizes and temperatures 
accessible to simulations are still insufficient to clearly solve several 
important questions~\cite{Young}. In particular the value of the critical 
temperature for the $d=2 \, \, \pm J$ Edwards-Anderson model is controversial.

Apart from the application of the roughness method to spin models, a central issue is 
the nature of the relationship between the correlations measured on the spin-state profiles
by the $\alpha$ or $\beta$ exponents and the traditional two-point correlation 
functions $C(r)$. We intend to examine this question more carefully in a future work.

\noindent{\large {\bf Acknowledgements}}

We thank Albens Atman and Jos\'e Guilherme Moreira for discussions. This work was supported by FAPEMIG and CNPq, Brazilian agencies.
 
\begin{figure}
\centerline{\epsfig{file=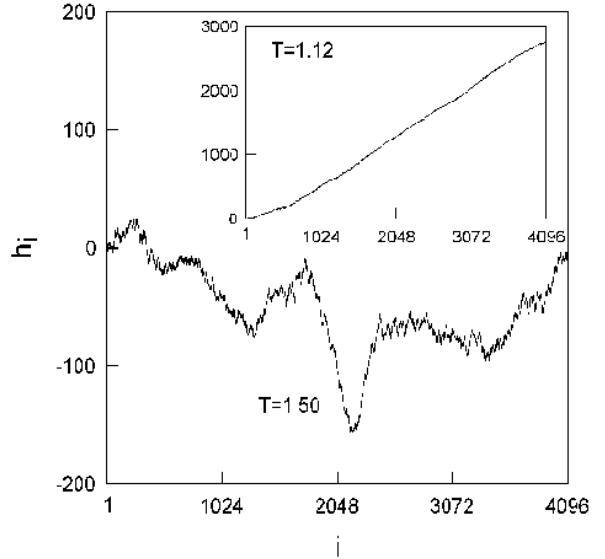,width=8cm,angle=0}}
\caption{Walk profiles obtained from equilibrium spin  configurations
of the Ising model ($q=2$) at $T=1.50>T_{c}^{(q=2)}$ and $T=1.12<T_{c}^{(q=2)}$ (inset).}
\label{fig1}
\end{figure}

\begin{figure}
\centerline{\epsfig{file=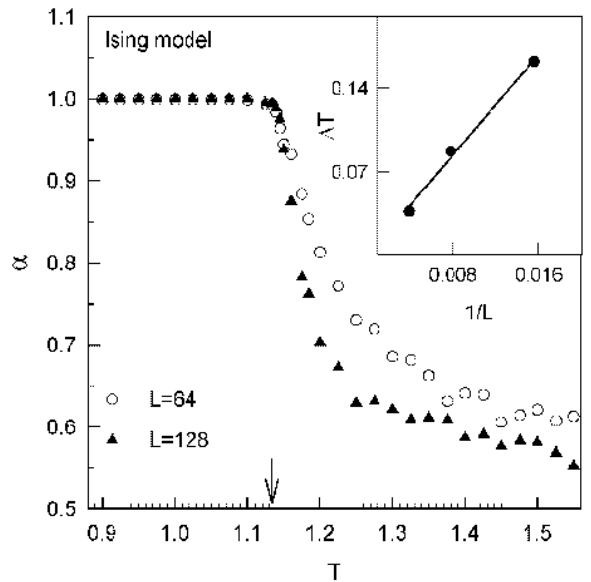,width=8cm,angle=0}}
\caption{The roughness exponent $\alpha$ as a function of temperature 
for the Ising model with system sizes $L=64$ and $L=128$. The arrow indicates
the static critical temperature $T_{c}^{(q=2)}$. The inset shows the behavior
of the width $\Delta T$ (see text) with the inverse of the system size $N$,
indicating that $\alpha (T)$ is a step function at the $N\rightarrow\infty$ limit.}
\label{fig2}
\end{figure}

\begin{figure}
\centerline{\epsfig{file=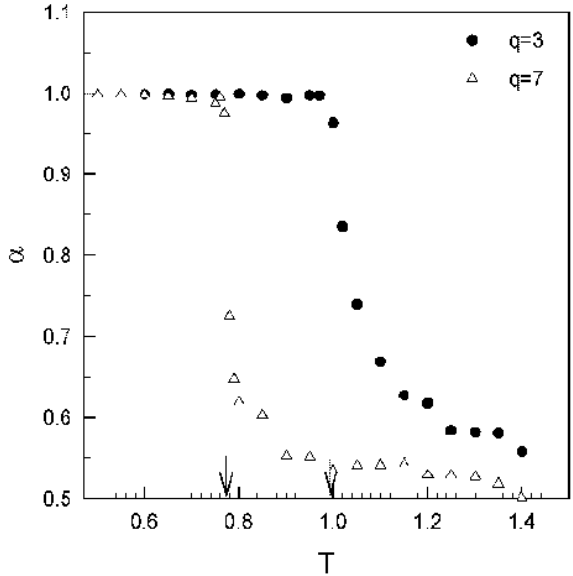,width=8cm,angle=0}}
\caption{The roughness exponent $\alpha$ as a function of temperature for 
the $q=3$ and $q=7$ Potts model with $L=128$. For $q=3$ the model exhibits 
a second order phase transition, and for $q=7$ a first order one.
The arrows indicate
the static critical temperatures $T_{c}^{(q=3)}$ and $T_{c}^{(q=7)}$. }
\label{fig3}
\end{figure}

\begin{figure}
\centerline{\epsfig{file=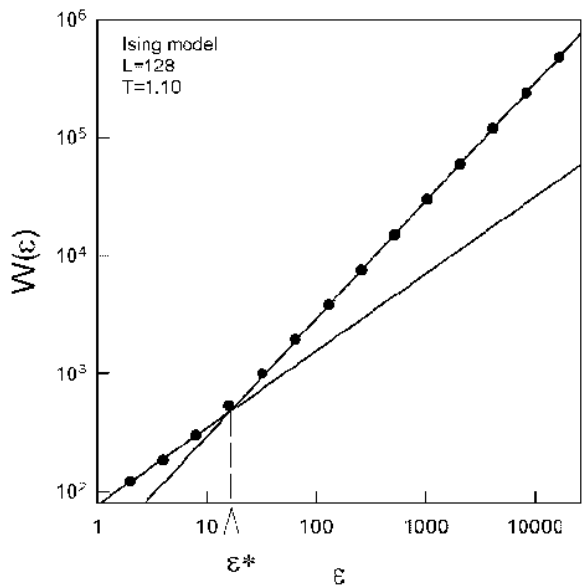,width=8cm,angle=0}}
\caption{A typical log-log plot of $W(\epsilon) \times \epsilon$ used to 
determine the roughness exponent $\alpha$ characterizing a profile obtained from an 
Ising spin equilibrium configuration for $T=1.10<T_{c}^{(q=2)}$. 
We can see clearly a  length scale $\epsilon^*$ which marks a change in the slope 
of the fitted straight lines.}
\label{fig4}
\end{figure}

\begin{figure}
\centerline{\epsfig{file=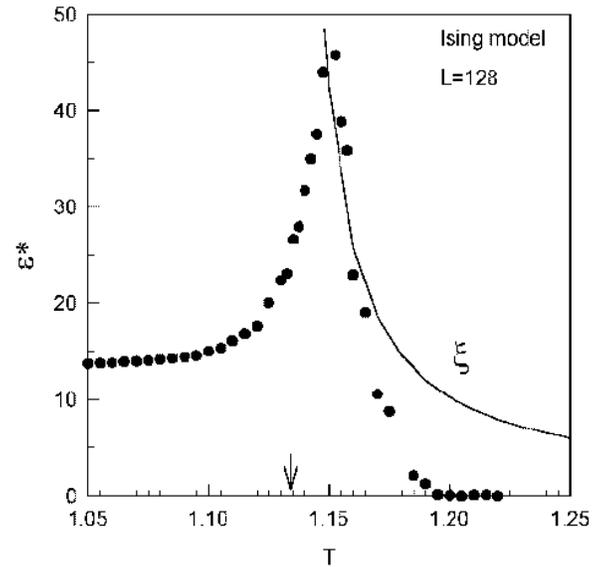,width=8cm,angle=0}}
\caption{The characteristic length $\epsilon^*$ as a function of temperature 
$T$ for the Ising model with system size $L=128$. 
$\epsilon^*$ exhibits a maximum around the critical temperature of the model.
The solid curve represents the exact behavior for the correlation length $\xi$
in the disordered phase.
The arrow indicates the static critical temperature $T_{c}^{(q=2)}$.}
\label{fig5}
\end{figure}

\begin{figure}
\centerline{\epsfig{file=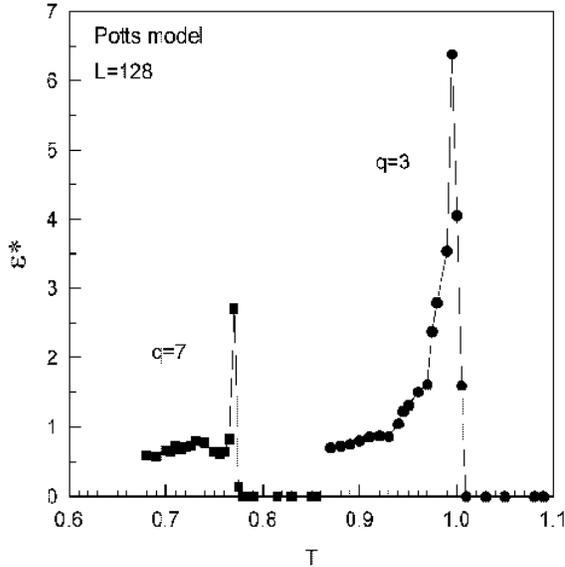,width=8cm,angle=0}}
\caption{The characteristic length $\epsilon^*$ as a function 
of temperature $T$ for the $q=3$ and $q=5$ Potts models with system size $L=128$.}
\label{fig6}

\end{figure}
\begin{figure}
\centerline{\epsfig{file=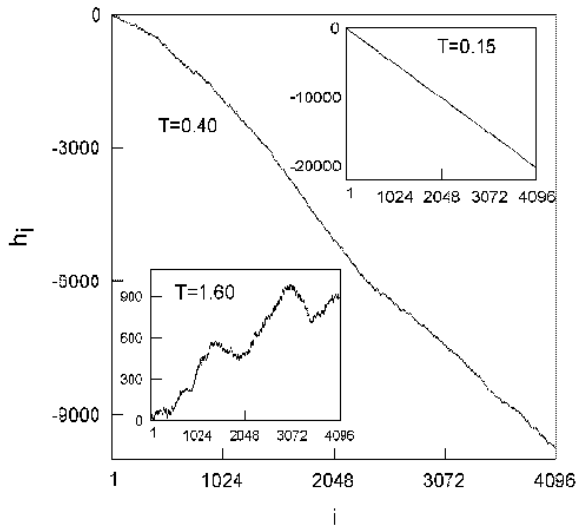,width=8cm,angle=0}}
\caption{Walk profiles obtained from equilibrium spin  configurations
of the $p=10$ clock model. $T=0.15$ is located in the ordered phase, $T=0.40$ is in the
spin-wave phase and $T=1.60$ is in the disordered one.}
\label{fig7}

\end{figure}
\begin{figure}
\centerline{\epsfig{file=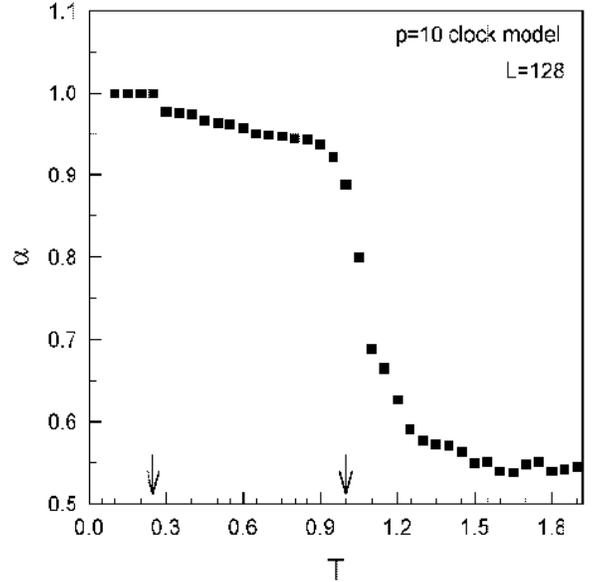,width=8cm,angle=0}}
\caption{The roughness exponent $\alpha$ as a function of temperature for the 
$p=10$ clock model with system size $L=128$. 
This model exhibits a Kosterlitz-Thouless type phase transition at $T\simeq 0.24$.}
\label{fig8}
\end{figure}

\begin{figure}
\centerline{\epsfig{file=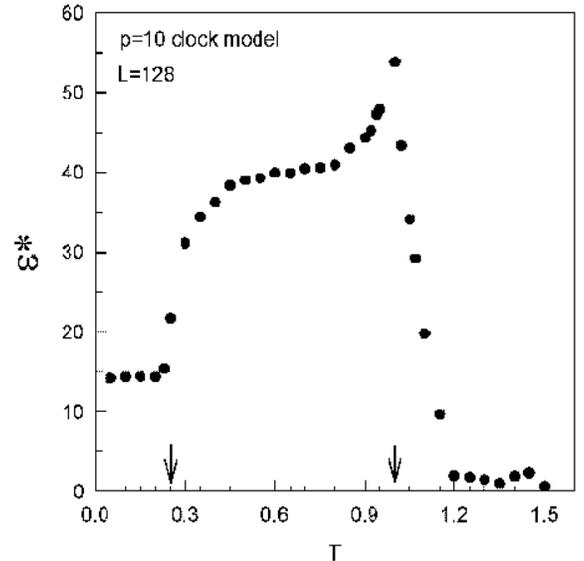,width=8cm,angle=0}}
\caption{The characteristic length $\epsilon^*$ as a function of temperature 
$T$ for the $p=10$ clock model with system size $L=128$. 
The arrows indicate
the static critical temperatures $T_{Ac}^{(p=10)}$ and $T_{Bc}^{(p=10)}$.}
\label{fig9}
\end{figure}

\end{multicols}
\widetext

\end{document}